\documentclass[12pt]{article}
\usepackage{amsmath}
\usepackage{epsfig}
\begin{document}
\title{Nucleon QCD sum rules in instanton vacuum
}
\author{E. G. Drukarev,~V. A. Sadovnikova\\
{\em National Research Center "Kurchatov Institute"}\\
{\em B. P. Konstantinov Petersburg Nuclear Physics Institute}\\
{\em Gatchina, St. Petersburg}}

\maketitle

\begin{abstract}
We calculate the polarization operator of the nucleon current in the instanton medium. The medium (QCD vacuum) is assumed to be a composition of the instantons of large and small sizes. The former are described in terms of the local scalar condensate, while the latter can be interpreted as the nonlocal scalar condensate. We solve the corresponding QCD sum rules equations and demonstrate that there is a solution with the value of the nucleon mass close to the physical one if the fraction of the small size instantons $w_s \approx 2/3$.
\end{abstract}

\section{Introduction}

The idea of the approach is to express the characteristics of the observed hadrons in terms of the
vacuum expectation values of the QCD operators, often referred to as the condensates. It was suggested in \cite{1}
for calculation of the characteristics of mesons. Later it was used for the nucleons \cite{2}.
It succeeded in calculation of the nucleon mass, anomalous magnetic moment, axial coupling constant, etc \cite{3}.

The QCD sum rules (SR) approach is based on the dispersion relation for the
function describing the propagation of the system which carries the quantum
numbers of the hadron. This function is usually referred to as the "polarization operator"  $\Pi(q)$,
with $q$ the four-momentum of the system. The dispersion relation in which we do not take care of the subtractions
\begin{equation}
\Pi(q^2)=\frac1\pi\int dk^2\frac{\mbox{Im}\Pi(k^2)}{k^2-q^2}
\label{0}
\end{equation}
is analyzed at large and negative values of $q^2$, since,
due to the asymptotic freedom of QCD the polarization operator can be calculated as a power series
of $q^{-2}$ at $q^2\to-\infty$. The coefficients of the expansion are the QCD condensates, such as the scalar quark condensate $\langle 0|\bar q(0)q(0)|0\rangle$, gluon condensate $\langle 0|G^{a\mu \nu}G^a_{\mu \nu}|0\rangle$, etc. This is known as the Operator Product Expansion (OPE) \cite{4}.
while the nonperturbative physics is contained in the
condensates. The typical values of condensate with the dimension $d=n$ is $\langle 0|O_n|0\rangle \sim (\pm 250 MeV)^n$. Thus  we expect the series $\Pi(q)=\sum_n\langle 0|O_n|0\rangle/(q^2)^n$ to converge at $-q^2 \sim 1~\,{\rm GeV}^2$.

The left hand side (LHS) of Eq.~(\ref{0}) is calculated as the OPE series. The imaginary part on the right hand side (RHS) describes the physical states with the baryon quantum number and charge equal to unity. These are the proton, described by the pole of ${\rm Im} \Pi(k^2)$, the cuts corresponding to the systems containing the proton and pions, etc.
The right-hand side  of Eq.~(\ref{0}) is usually approximated by the "pole+continuum" model
\cite{1,2} in which the lowest lying pole is written down exactly,
while the higher states are described by continuum.
The main aim is to obtain the value of the nucleon mass.

The polarization operator can be written as
\begin{equation}
\Pi(q^2)=i\int d^4xe^{i(q\cdot x)} \langle 0|T[j(x)\bar j(0)]|0 \rangle\,,
\label{6}
\end{equation}
where $j(x)$ the local operator with the proton quantum numbers,
often referred to as ``current". It is a composition of the quark operators.
Thus the integrand on the RHS of Eq.~(\ref{6}) contains the nonlocal expectation values
$\langle 0|\bar q(0)q(x)|0\rangle$.

Note that the product $\bar q(0)q(x)$ is not gauge invariant. This expression makes sense if we define $q(x)$ as the Taylor expansion near the point $x=0$, i.e.
\begin{equation}
q(x)=\,\left(1+x^{\mu}D_{\mu}+\frac{x^{\mu}x^{\nu}}{2}D_{\mu}D_{\nu}+...\right) q(0)\,,
\label{300a}
\end{equation}
with $D^{\mu}$ standing for covariant derivatives. Thus the condensate $\langle 0|\bar q(0)q(x)|0\rangle$ can be expressed in terms of the set of new nonlocal condensates, such as$\langle 0|\bar q(0)D^2q(0)|0\rangle$.

In this approach the QCD condensates are considered as phenomenological parameters. Extractions of their  values from experimental data, supported by certain theoretical ideas does not always lead to unique conclusions.

One usually applies the Borel transform which converts the functions of $q^2$ to the functions
of the Borel mass $M^2$. Note also that the Borel transform removes the divergent contributions caused by behavior of the integrand in the integral on the RHS of Eq.~(\ref{6}) at the lower limit. An important assumption is that there
is an interval of the values of $M^2$ where the two sides of the SR have a good
overlap, approximating also the true functions. This interval is in the region of $1\,{\rm GeV}^2$. Thus actually
one tries to expand the OPE from the high momentum region to the region of $ |q^2| \sim 1\,{\rm GeV}^2$.

The nowadays point of view is (see, e.g., \cite{5}) that the $\bar qq$ pairs which form the scalar quark condensates are produced by the strong gluon fields (instantons) which compose the QCD vacuum. We try to write the QCD sum rules in terms of the quarks propagating in instanton vacuum. The common belief is that the instantons produce mainly the $\bar qq$ pairs which compose the condensate ("zero mode"). Some approximations are usually done for the other quark states in the instanton field ("nonzero modes"). In the present paper we compose and solve the QCD sum rules based on this structure of the QCD vacuum.

A typical size of instanton is $\rho \approx (600 {\rm MeV})^{-1}$, which is not much larger than a typical inverse Borel mass
$1/M \approx (1\,{\rm GeV})^{-1}$. Thus the instanton effects depend on the parameter $\rho M \sim 1$, and we can not employ the OPE. In other words, we can not neglect the inhomogeneous structure of the instanton field at the distances
of the order $1/M$ and should take into account that the instanton creates a nonlocal quark condensate  $\langle 0|\bar q(0)q(x)|0\rangle$. Thus one of the consequences of instanton presentation  is that we include (although not in a straightforward way) the nonlocality of the scalar condensate.

The instantons have various sizes
$\rho$, and their distribution $n(\rho)$ is presented, i.e., in \cite{6}. We approximate the continuous distribution
$n(\rho)$ as a sum of two discrete values corresponding to "large size" and "small size" instantons $n(\rho)=n_{\ell}\delta (\rho-\rho_{\ell})+n_s\delta(\rho-\rho_s)$. Thus we can write
 \begin{equation}
\langle 0|\bar q(0)q(x)|0\rangle=\langle 0|\bar q(0)q(x)|0\rangle_{\ell}+\langle 0|\bar q(0)q(x)|0\rangle_s\,.
\label{6a0}
\end{equation}

The size $\rho_{\ell}$ is assumed to be much larger than the inverse value of the typical Borel mass, and we can neglect the nonlocality of the scalar condensate on this scale. The typical small instanton size is $\rho_s \sim 1/(600\,{\rm MeV})$, and the nonlocality should be included.
Thus we put
\begin{equation}
\langle 0|\bar q(0)q(x)|0\rangle=\langle 0|\bar q(0)q(0)|0\rangle_{\ell}+\langle 0|\bar q(0)q(x)|0\rangle_s\,.
\label{6b}
\end{equation}
Introducing the fraction of small instantons $w_s$ we can write
\begin{equation}
\langle 0|\bar q(0)q(x)|0\rangle=\langle 0|\bar q(0)q(0)|0\rangle w_{\ell}+\langle 0|\bar q(0)q(x)|0\rangle w_s\,;
\quad w_s+w_{\ell}=1.
\label{6a}
\end{equation}

We treat the quarks in the field of small instantons, following the approach developed in \cite{7,8}.
In this approach the zero mode contribution to the propagator of the light quark carrying momentum $p$ contains  a chirality-flipping factor $m(p)$. The nonzero mode contribution is described by the free quark propagator. This leads to significant changes in the structure of the LHS of the sum rules. Additional approximation is that the quarks of polarization operator can interact only with one small instanton.  This happens since  we consider the momenta $|q^2| \sim 1\,{\rm GeV}^2$.

The OPE can not be employed for the calculation of the LHS of the SR, which is obtained as a function of the Borel mass $M^2$. For the chirality flipping sum rule the function of $M^2$ on the LHS can be viewed as
coming from the nonlocality of the scalar quark condensate.
The contribution of the four-quark condensate presented in instanton picture provides now a much smaller contribution since one instanton can produce only one $\bar qq$ pair of the fixed flavor. On the other hand some of the condensates which contribute to the OPE SR are not accounted for in our model for the nonzero modes contribution. Thus we can not speak about "inclusion of instantons to the QCD sum rules".
The languages of local condensates and instantons are rather complimentary to each other.

We calculate the polarization operator $\Pi(q)$ in the instanton vacuum and analyze the corresponding SR.
We demonstrate that the SR have a solution with the value of the nucleon mass not far from the physical one for
all $w_s < 0.6-0.7$. At $w_s \approx 2/3 $ the value of the nucleon mass is $m_N \approx 1$ GeV. Comparing with the SR in terms of condensates we found that the consistency between the LHS and RHS of the SR improved. At the conventional values of the quark condensate $\langle 0|\bar q(0)q(x)|0\rangle \approx (-250 $MeV$)^3$ the value of the threshold does not change much, while that of nucleon residue becomes noticeably smaller.
At larger values of $w_s$ the SR have only an unphysical solution with the contribution of the continuum exceeding much that of the nucleon pole.
In Sec.~2 we recall the main features of the nucleon  SR in terms of condensates. In Sec.~3 we calculate the polarization operator in instanton vacuum. In Sec.~4 we solve the SR equations. We discuss the results in Sec.~5.

\section{QCD sum rules in terms of condensates}

In the case of nucleon (we consider the proton) the polarization operator takes the form
\begin{equation}
\Pi(q)=\hat q\Pi^q(q^2)+I\Pi^I(q^2)\,,
\label{1}
\end{equation}
with $q$ the four-momentum of the system, $\hat q=q_{\mu}\gamma^{\mu}$, $I$ is the unit matrix. The first and the second terms on the RHS correspond to the chirality conserving and the chirality flipping contributions, correspondingly. The dispersion relations are
 \begin{equation}
\Pi^i(q^2)=\frac1\pi\int dk^2\frac{\mbox{Im}\Pi^i(k^2)}{k^2-q^2}\,;
 \quad i=q,I .
\label{2}
\end{equation}
As we said earlier, we do not take care of the subtractions.

To calculate the polarization operator defined by Eq.~(\ref{6}) we must clarify the form of the current $j(x)$.
 It is not unique. One can write
\begin{equation}
j(t,x)=j_1(x)+tj_2(x)\,,
\label{7}
\end{equation}
with
$$j_1(x)=(u^T_a(x)Cd_b(x))\gamma_5u_c(x) \varepsilon^{abc}, \quad j_2(x)=(u^T_a(x)C\gamma_5d_b(x))u_c(x) \varepsilon^{abc},$$
while $t$ is an arbitrary coefficient. Following \cite{9} we use the current determined by Eq.~(\ref{7})
with $t=-1$, which can be written (up to a factor $1/2$) as
\begin{equation}
j(x)=(u^T_a(x)C\gamma_{\mu}u_b(x))\gamma_5 \gamma^{\mu}d_c(x) \varepsilon^{abc}\,.
\label{8} \end{equation}
This current is often used in the QCD SR calculations. One of the strong points of the choice is that it makes the domination of the lowest pole over the higher states on the RHS of Eq.~(\ref{2}) more pronounced. We use only this current in the present paper.

The LHS of Eq.~(\ref{2}) can be written as
\begin{equation}
\Pi^{q~OPE}(q^2)=\sum_{n=0}A_n(q^2); \quad \Pi^{I~OPE}(q^2)=\sum_{n=3} B_n(q^2)
\label{3a}
\end{equation}
where the lower index $n$ is the dimension of the corresponding QCD condensate ($A_0$ stands for the three-quark loop).
The most important terms for $n\leq 8$ were obtained earlier \cite{2,3}.
For the chirality conserving structure they are
\begin{equation}
A_0=\frac{-Q^4\ln{Q^2}}{64\pi^4}; \quad A_4=\frac{-b\ln{Q^2}}{128\pi^4}; \quad A_6=\frac{1}{24\pi^4}\frac{a^2}{Q^2};
\quad A_8=-\frac{1}{6\pi^4}\frac{m^2_0a^2}{Q^4}\,.
\label{3c}
\end{equation}
Here $Q^2=-q^2>0$, while $a$ and $b$ are the scalar and gluon condensates multiplied by certain numerical factors
\begin{equation}
a=-(2\pi)^2\langle 0|\bar q q|0\rangle  ; \quad b=(2\pi)^2\langle 0|\frac{\alpha_s}{\pi}G^{a\mu \nu}G^a_{\mu \nu}|0\rangle\,,
\label{3d}
\end{equation}
while
\begin{equation}
m^2_0 \equiv \frac{\langle 0|\bar q\sigma_{\mu\nu}{\cal G}_{\mu\nu}q|0\rangle}{\langle 0|\bar q q|0\rangle};
\quad {\cal G}_{\mu\nu}=\frac{\alpha_s}{\pi}\sum_hG^h_{\mu\nu}\lambda^h /2
\label{3dd}
\end{equation}
We shall discuss the value of the $m_0^2$ in Sec.~5.
For the chirality flipping structure we find
\begin{equation}
B_3=\frac{aQ^2\ln{Q^2}}{16\pi^4}; \quad B_5=0\,.
\label{3ddd}
\end{equation}
The leading contribution to the chirality conserving structure $ A_0$ is the loop containing three free quarks. The leading contribution to the chirality-odd structure $B_3$ is proportional to the scalar quark condensate. Here the free $u$ quarks form a loop, while the $d$ quarks are exchanged with the vacuum condensate - see Fig.~1.

Let us, however, pay attention to the latter equality $B_5=0$. There are indeed two contributions of dimension
$d=5$. Thus we can write $B_5=B_5^a+B_5^b$. The term $B_5^a$ comes from the Taylor expansion of the product
$\bar d(0)d(x)$ and is proportional to the condensate $\langle 0|\bar d(0)D^2d(0)|0\rangle$. In this case
the $u$ quarks are described by free propagators which are diagonal in color variables. However the product
of the operators $d^a_{\alpha}\bar d^b_{\beta}G^h_{\mu\nu}$ provide the  contribution to the propagator of $d$ quark which is proportional to the product $\langle 0|\bar q {\cal G}_{\mu\nu}\sigma_{\mu\nu}q|0\rangle\cdot\sigma_{\alpha\beta}\lambda^h_{ab}/2$. The contribution to polarization operator $B^b_5$ is thus
proportional to the condensate $\langle 0|\bar q{\cal G}_{\mu\nu}\sigma_{\mu\nu}q|0\rangle$, and the propagator
of one of the $u$ quarks of polarization operator should include interaction with this gluon field (and can not be treated as a free one). Due to equation of motion
$$ (D^2-\frac{1}{2}\sigma_{\mu\nu}{\cal G}_{\mu\nu})q=m^2_qq,$$
with $m_q$ standing for the current mass of the quark, one finds that for the massless quark
\begin{equation}
\langle 0|\bar d(0)D^2d(0)|0\rangle=\frac{1}{2}\langle 0|\bar q\sigma_{\mu\nu}{\cal G}_{\mu\nu}q|0\rangle\,.
\label{4ddd}
\end{equation}
Thus the contributions $B_5^a$ and $B_5^b$  can be expressed in terms of the same condensate.
Direct calculation \cite{I1} demonstrates that $B_5^a+B^b_5=0$. Note that this cancelation takes place only for the current (\ref{8}). If one employs the current (\ref{7}) with $t \neq -1$, the contribution $B_5 \neq 0$.

Actually one usually considers the SR for operators ${\cal P}^i(M^2)=32\pi^4{\cal B}\Pi^{i\,OPE}(q^2)$ with ${\cal B}$
the operator of Borel transform. The factor $32\pi^4$ is
introduced in order to deal with the values of the order of unity (in
GeV units). After the Borel transform we write (\ref{3a}) as
\begin{equation}
{\cal P}^q(M^2)=\sum_{n=0}A'_n(M^2); \quad {\cal P}^i(M^2)=\sum_{n=3} B'_n(M^2);
\quad A'_n(M^2)=32\pi^4{\cal B}A_n(q^2)\,.
\label{3b}
\end{equation}
$$ B'_n(M^2)=32\pi^4{\cal B}B_n(q^2).$$
Here we present the most important terms
\begin{equation}
A'_0(M^2)=M^6; \quad A'_4(M^2)=\frac{bM^2}{4}; \quad
A'_6=\frac43 a^2;\quad
B'_3(M^2)=2aM^4\,.
\label{22a}
\end{equation}
The Borel transformed SR (\ref{2}) can be written now as
\begin{equation}
{\cal P}^i(M^2)={\cal F}_p^{i}(M^2)+{\cal F}_c^{i}(M^2)\,,
\label{3g}
\end{equation}
where the two terms on the RHS are the contributions of the nucleon pole with the mass $m_N$ and that of the continuum
\begin{equation}
{\cal F}_p^{i}(M^2)=\xi_i\lambda^2e^{-m_N^2/M^2};
\quad {\cal F}_c^{i}(M^2)=\int_{W^2}^{\infty}dk^2e^{-k^2/M^2}\Delta[{\cal B}\Pi_1(k^2)]
\label{3h}
\end{equation}
to the RHS of the Borel transformed Eq.~(\ref{2}). Here $\lambda$ is the residue at the nucleon pole (multiplied by $32\pi^4$), $W^2$ is the continuum threshold;
$\xi_q=1$, $\xi_I=m_N$.

The conventional form of the SR is
\begin{equation}
{\cal L}^q(M^2, W^2)=R^q(M^2)\,,
\label{4}
\end{equation}
and
\begin{equation}
{\cal L}^I(M^2, W^2)=R^I(M^2)\,.
\label{4a}
\end{equation}
Here ${\cal L}^i$ and $R^i$ are the Borel transforms of the LHS and of the RHS of Eqs.~(\ref{2}), correspondingly
\begin{equation}
R^q(M^2)=\lambda^2e^{-m_N^2/M^2}; \quad R^I(M^2)=m_N\lambda^2e^{-m_N^2/M^2}\,,
\label{5}
\end{equation}
with $\lambda^2=32\pi^4\lambda_N^2$. The contribution of continuum is moved to the LHS of
Eqs.~(\ref{4}, \ref{4a}) which can be written as
\begin{equation}
{\cal L}^q=\sum_{n=0} \tilde A_n(M^2, W^2); \quad {\cal L}^I=\sum_{n=3}\tilde B_n(M^2, W^2)\,,
\label{5a}
\end{equation}
-see Eq.~(\ref{3b}). Here

\begin{equation}
\tilde A_0=\frac{M^6E_2(\gamma)}{L(M^2)};\quad
\tilde A_4=\frac{bM^2E_0
(\gamma)}{4L(M^2)};
\label{22}
\end{equation}
$$\tilde A_6=\frac43 a^2L; \quad
\tilde B_3=2aM^4E_1(\gamma); \quad \gamma=\frac{W^2}{M^2}\,,$$
with
\begin{equation}
E_0(\gamma)=1-e^{-\gamma}, \quad E_1(\gamma)=1-(1+\gamma)e^{-\gamma}, \quad E_2(\gamma)=1-(1+\gamma+\gamma^2/2)e^{-\gamma}\,.
\label{23} \end{equation}
The factor
\begin{equation}
L(M^2)=\Big(\frac{\ln M^2/\Lambda^2}{\ln \mu^2/\Lambda^2}\Big)^{4/9}
\label{3100} \end{equation}
includes the most important radiative corrections of the order $\alpha_s\ln{Q^2}$ (LLA). These contributions were summed to all orders of $(\alpha_s\ln{Q^2})^n$.
In Eq.~(\ref{3100}) $\Lambda=\Lambda_{QCD}$ is the QCD scale, while $\mu$ is the
normalization point, the standard choice is $\mu=0.5\,$GeV.

The position of the nucleon pole $m_N$, its residue $\lambda^2$ and the continuum threshold $W^2$ are the unknowns of the SR equations (\ref{4}) and (\ref{4a}). The nucleon sum rules equations (\ref{4}) and (\ref{4a}) are usually solved at $M^2 \sim 1\,{\rm GeV}^2$, namely
\begin{equation}
0.8 \mbox{GeV}^2 \leq M^2 \leq 1.4\,\mbox{GeV}^2\,.
\label{230}
\end{equation}
The interval of the values of $M^2$ where the SR are true is usually referred to as "duality interval".

After inclusion of several condensates of the higher dimension and of the lowest order radiative corrections beyond the leading logarithmic approximation \cite{10}
the SR provide solution (for $\Lambda_{QCD}=230$ MeV) $m_N=928$ MeV, $\lambda^2=2.36\,{\rm GeV}^6$, $W^2=2.13\,{\rm GeV}^2$.

\section{QCD sum rules in terms of instantons }

The contribution  of the large size instantons  can be included by changing of the
quark condensate $a$ to $aw_{\ell}$ in the OPE terms - see Eq.~(\ref{6a}). The analysis of the contribution of the small size instantons is  more complicated.

\subsection{One instanton approximation}
We must calculate polarization operator describing the quarks by their Green functions in the medium of the small size instantons. We employ the approach developed in \cite{7,8} by averaging of the propagator in the single-instanton field over the system of instantons and antiinstantons.
A closed form for the propagator in instanton medium obtained in \cite{7,8} is \footnote{In Sec.~3 we use the Euclidean metric which is more convenient for description of the instanton physics.}
\begin{equation}
S_{ab}(p)=\frac{\hat p+im(p)}{p^2+m^2(p)}\delta_{ab}\,,
\label{31a}
\end{equation}
with the effective dynamical mass $m(p)$ found in \cite{7,8} as a function of the small instanton size $\rho$ and the distance between the instantons $R$. Actually $R \approx 3\rho$.

The average distance between the instantons is estimated to be $R \approx 1$~fm. It is much larger than the
inverse Borel mass $1/M \sim 0.2$~fm. Thus the size of the system described by the polarization operator is much smaller
than $R$ and can accommodate only one instanton. This leads to several consequences.

We can write the quark propagator rather as
\begin{equation}
S_{ab}(p)=S_Z+S_{NZ}; \quad S_Z=\frac{im(p)}{p^2}\delta_{ab}; \quad S_{NZ}=\frac{\hat p}{p^2}\delta_{ab}\,.
\label{31}
\end{equation}
Here $S_Z$ is the zero-mode contribution. The sum of all the nonzero-mode contributions $S_{NZ}$ is approximated thus by the free propagator of the massless quark.

The general relation between the quark condensate and the propagator
\begin{equation}
\langle 0|\bar q(0)q(0)|0\rangle_s=i\int\frac{d^4p}{(2\pi)^4}{\rm Tr} S(p) = -4N_c \int \frac{d^4p}{(2\pi)^4} \frac {m(p)}{p^2}\,,
\label{8a} \end{equation}
where we kept the Minkowsky metric for the quark operators in the vacuum expectation values, can be written now as
\begin{equation}
a_s=6\int_0^{\infty}dp\,p\,m(p); \quad a_s=-(2\pi)^2\langle 0|\bar q q|0\rangle_s\,.
\label{33A} \end{equation}

The structure of the LHS of the sum rules differs from that in the condensate presentation. The leading contribution $A_0$ to the operator $\Pi^q$ remained unchanged. However, there is no contribution of two zero-mode $u$ quarks (this was the four-quark condensate in the "condensate language"), since only one $u$ quark can be placed in the zero mode
of the field of the small size instanton.

In the chirality flipping structure $\Pi^I$ we describe the $d$ quark by the propagator $S_Z$ given by Eq.~(\ref{31}).
The Borel transformed contribution ${\cal B}\Pi^I(M^2)$ depends on the parameter $\rho_s^2M^2 \sim 1$ and can not be presented as $1/M^2$ series. In the condensate language this means that it includes the nonlocal scalar quark condensate.

Note that our presentation of the propagator $S_{NZ}$ means that we did not pick some of contributions which were present in the condensate presentation. In the terms $A_4$ and $B_5^b$ the propagator of one of the $u$ quarks should
include the influence of the gluon field. On the other hand, the term $B_5^a$ described the nonlocality of the scalar condensate in the lowest order of $x^2$ expansion.

We include the leading logarithmic corrections to the current $j$ and to the large size condensates. We do not include them for the contributions of small instantons, since this is beyond the accuracy of the approach.

\subsection{Calculation of polarization operator}

As we said earlier, the leading contribution  $A_0$ to the $\hat Q$ structure
remains unchanged. The contribution to the chirality flipping structure is now
\begin{equation}
\Pi_1^I(q^2)=2a(1-w_{s})Q^2\ln{Q^2}+Y_s\,,
\label{34}
\end{equation}
where the two terms are the contributions of the large size and small size instantons, correspondingly. The last one can be written as $Y_s=32\pi^4X_s$, with
\begin{equation}
X_s=12\int\frac{d^4p}{(2\pi)^4}\gamma_{\mu}\frac{m(p)}{p^2}\gamma_{\nu}T_{\mu\nu}(Q-p)\,,
\label{48}
\end{equation}
while
\begin{equation}
T_{\mu\nu}(Q-p)=\int d^4x e^{-i(Q-p,x)}{\rm Tr}[t_{\mu\nu}(x)]\,,
\label{35}
\end{equation}
with
\begin{equation}
t_{\mu\nu}(x)=\gamma_{\mu}G_0(x)\gamma_{\nu}G_0(x)\,.
\label{36}
\end{equation}
Here
\begin{equation}
G_0(x)=-\frac{1}{2\pi^2}\frac{\hat x}{x^4}
\label{37}
\end{equation}
is the Fourier transform of the propagator $S_{NZ}$ determined by Eq.~(30).
Note that putting $p=0$ in the factor $ T_{\mu\nu}(Q-p)$ on the LHS of Eq.~(\ref{48}) we would obtain
\begin{equation}
X_s=\frac{3Q^2\ln{Q^2}}{8\pi^4}\int_0^{\infty} dp\,p\,m(p)=B_3(Q^2)\,,
\label{50}
\end{equation}
with $B_3(Q^2)$ defined by Eq.~(\ref{3c}) and $a$ replaced by $a_s$.
Thus in the limit $Q^2 \rightarrow \infty$ we obtain the lowest OPE term.
We can view calculation of the contribution given by Eq.~(\ref{48}) as inclusion of nonlocality in the scalar quark condensate.

The four-quark contribution can emerge only if one of the $\bar u u$ pairs comes from the small size instantons, while the other one originates from large size ones. Following the previous discussion we can write the contribution to polarization operator as
\begin{equation}
A_6=\frac{4a^2w_s(1-w_s)}{\pi^2}\int \frac{d^4p}{(2\pi)^4}\frac{m(p)}{p^2}
\frac{\hat Q-\hat p}{(Q-p)^2}\,.
\label{63}
\end{equation}
Here the lower index $6$ shows that in the limit $M^2 \rightarrow \infty$ the contribution turns to the OPE term
$\tilde A_6$ determined by Eq.~(\ref{22a}) multiplied by $2w_s(1-w_s)$.
The set of diagrams included in the SR is shown in Fig.~2.

To obtain results in analytical form we parameterize the dynamical quark mass caused by the small size instantons
\begin{equation}
m(p)=\frac{{\cal A}}{(p^2+\eta^2)^3}\,,
\label{53}
\end{equation}
with ${\cal A}$ and $\eta$ the adjusting parameters. The power of denominator insures the proper behavior
$m(p) \sim p^{-6}$ at $p \rightarrow \infty$ \cite{7}.
Now Eq.~(\ref{33A}) can be written as
\begin{equation}
a_s=\frac{3{\cal A}}{2\eta^4}\,.
\label{54}
\end{equation}

Calculating the tensor $T_{\mu\nu}$ we present
\begin{equation}
X_s=\frac{3}{\pi^2}\int \frac{d^4p}{(2\pi)^4} \frac{{\cal A}}{p^2(p^2+\eta^2)^3}(Q-p)^2\ln{(Q-p)^2}\,.
\label{49}
\end{equation}

Further details of calculations are presented in Appendix. We find for the Borel transformed
contribution
\begin{equation}
B'(M^2)=2a_{\ell}M^4+2a_sM^4F(\beta); \quad F(\beta)=\frac{1}{3}\Big(\frac{2(1-e^{-\beta})}{\beta}+e^{-\beta}(1-\beta) +\beta^2{\cal E}(\beta)\Big)\,;
\label{55}
\end{equation}
$$\beta=\eta^2/M^2.$$
Here
\begin{equation}
{\cal E}(\beta)=\int_{\beta}^{\infty}dt\frac{e^{-t}}{t}\,.
\label{56}
\end{equation}
In literature our function ${\cal E}$ is usually denoted as $E_1$. We avoid this notation since in QCD SR
publications the notation $E_1$ has another meaning - see Eq.~(\ref{23}).

Combining Eq.~(\ref{54}) with the relation $m(0)={\cal A}/\eta^6$ coming from Eq.~(\ref{53}) we find that $\eta^2=2a_s/3m(0)$. It was demonstrated in \cite{7,8} that $a_s \sim R^{-2}\rho^{-1}$, while $m(0)\sim
R^{-2}\rho$. Thus $\eta^2$ depends only on $\rho$, and  $\eta^2=1.26\,{\rm GeV}^2$ at $\rho=0.33$~fm.
In the duality interval (\ref{230}) $0.9 \leq \beta \leq 1.6$.
The function $F(\beta)$ is shown in Fig.~3a.
The dependence of the function $F$ on $M^2$ for $\eta^2=1.26\,{\rm GeV}^2$ is shown in Fig.~3b.
As expected, in the limit $M^2 \rightarrow \infty$ we find $B=B'_3$ with the latter defined by Eq.~(\ref{22a}).

Similar calculation provides
\begin{equation}
A'_6=\frac{8}{3}a^2w_s(1-w_{s})\frac{1-e^{-\beta}}{\beta}\,.
\label{63a}
\end{equation}

\subsection{Connection with the OPE}
It is reasonable to try to establish connection with the OPE approach. We write Eq.~(\ref{55}) as
\begin{equation}
B'_3(M^2)=2M^4a(M^2)\,,
\label{80cc}
\end{equation}
where
\begin{equation}
a(M^2) = a\left(1-w_s+w_sF(\frac{\eta^2}{M^2})\right)\,,
\label{80ca}
\end{equation}
while $F$ is defined by Eq.~(\ref{55}). We have $a(M^2) \rightarrow a$ at $M^2 \rightarrow \infty $.
Now we define
\begin{equation}
K(M^2)=\frac{a(M^2)}{a}\,,
\label{80c}
\end{equation}
and try to find the function $K(M^2)$ as a power series in $1/M^2$:
\begin{equation}
K(M^2)=1+\sum_{n=1}^{N}C_n/M^{2n}\,,
\label{80d}
\end{equation}

If the second term on the RHS can be approximated by one or two terms,
such presentation can be related to the expansion
of the expectation value
$\langle 0|\bar q(0)q(x)|0\rangle$ in powers on  $x^2$.
We can write the polarization operator $\Pi_I$ as
\begin{equation}
\Pi_I(q^2)=\frac{2}{\pi^4}\int \frac{d^4x}{x^6}f(x^2)e^{iqx}\,,
\label{80g}
\end{equation}
with $f(x^2)=\langle 0|\bar q(0)q(x)|0\rangle$.
Assuming that $f(x)$ can be approximated by the polynomial
(recall that we are in Euclidean metric)
\begin{equation}
f(x)=f(0)(1+c_1x^2+c_2x^4)\,,
\label{81xx}
\end{equation}
we find
\begin{equation}
B'_3(M^2) = 2M^4f(0)\left(1-\frac{8c_1}{M^2}+\frac{32c_2}{M^4}\right)\,,
\label{81yy}
\end{equation}
and thus
\begin{equation}
c_1=\frac{C_1}{8}; \quad c_2=\frac{C_2}{32}\,.
\label{81yyy}
\end{equation}
Note that the RHS of Eq.~(\ref{81xx}) can not be treated as the lowest terms of the Taylor expansion. The terms $x^{2n}$ with $n \geq 3$ provide the integrals which are divergent on the upper limit and can not be eliminated by the Borel transform.

For the medium consisting solely of the small size instantons, i.e. for $w_s=1$ we find
in the interval $0.8\,{\rm GeV}^2 \leq M^2\leq 1.4\,{\rm GeV}^2$ - see Eq.~(\ref{230})
\begin{equation}
C_1=-1.23\mbox{ GeV}^2; \quad C_2=0.54\mbox{ GeV}^4\,,
\label{82yyy}
\end{equation}
and thus $c_1=-0.16\,{\rm GeV}^2$, $c_2=0.017\,{\rm GeV}^4$.
The accuracy of the parametrization is illustrated by Fig.~4.
This point was discussed in more details in \cite{M1}.

\section{Solution of the sum rules equations}

Now we return to the Minkowsky metric and analyze  Eqs.~(\ref{4}) and (\ref{4a}) with
\begin{equation}
{\cal L}^q=\tilde A_0(M^2, W^2)+ \tilde A_6(M^2)  ; \quad {\cal L}^I=\tilde B(M^2, W^2)\,.
\label{67a}
\end{equation}
Here $\tilde A_0(M^2, W^2)$ is given by Eq.~(\ref{22}), $\tilde A_6=A'_6$ is presented by Eq.~(\ref{63a}), while
\begin{equation}
\tilde B(M^2,W^2)=2a_{\ell}M^4E_2(\gamma)+2a_sM^4\Phi(M^2, W^2)\,;
\label{67}
\end{equation}
$$\Phi(M^2, W^2)=\frac{1}{3}\Big(\frac{2}{\beta}(1-e^{-\beta})+e^{-\beta}(1-\beta)-e^{-\gamma}(1-\beta+\gamma) +\beta^2({\cal E}(\beta)- {\cal E}(\gamma)) \Big)\,.$$
The functions $E_i (i=0,1,2)$ are determined by Eq.~(\ref{23}).

\subsection{Lack of solution at $w_s=1$}

One can guess immediately that there is no solution for $w_s=1$. In other words, the nucleon sum rules do not work in the word which consists only of small size instantons. Indeed, if the values $m_N, \lambda^2, W^2$ compose a solution, one
should obtain
\begin{equation}
\kappa(M^2)\equiv \frac{{\cal L}^I(M^2,W^2)}{{\cal L}^q(M^2,W^2)} \approx const=m_N
\label{167}
\end{equation}
since the contribution of the continuum should not be too large, we should expect
\begin{equation}
\frac{{\cal L}^I(M^2)}{{\cal L}^q(M^2)} \approx const \approx m_N\,,
\label{167a}
\end{equation}
where we put ${\cal L}^i(M^2)={\cal L}^i(M^2,W^2 \rightarrow \infty)$.

For $w_s=1$ Eq.~(\ref{167a}) takes the form
\begin{equation}
\kappa(M^2)=\frac{2aF(\eta^2/M^2)}{M^2}\,.
\label{167b}
\end{equation}

Employing the dependence of the function $F$ on $M^2$ for $\eta^2=1.26\,{\rm GeV}^2$ presented in Fig.~3b,
one can see that the values of $\kappa$ varies  between $2a\cdot 0.36/{\rm GeV}^2$ and $2a\cdot 0.27/{\rm GeV}^2$
in the interval (\ref{230}) of variation of $M^2$. For the distance $R=1$~fm between the small size instantons
$a=0.59\,{\rm GeV}^3$ \cite{7,8}.
Thus we obtained $m _N\approx 0.35$ GeV.

The unrealistic value of the nucleon mass obtained in such a way is, however, not the main problem.
Let us try to find the value of $\lambda^2$ employing Eq.~(21). We obtain $M^6\,e^{m^2_N/M^2} = \lambda^2$.
However the LHS of this equality changes by a factor of $6$ in the duality interval (\ref{230}).
Thus it can be satisfied only if the contribution of the continuum changes its LHS strongly.
Hence, we came to an unphysical solution of the sum rules \cite{11a}. As we shall see below, a more detailed analysis confirms this conclusion.

\subsection{Dependence of the solutions on the fraction of small size instantons}
The functions ${\cal L}^q$ and ${\cal L}^I$ depend explicitly on the scalar condensate $a$, on its fraction caused by the
instantons of the small size $a_s=aw_s$ and on the parameter $\eta^2$. On the other hand, the medium of the small instantons is determined by their average size $\rho$ and the distance between the instantons $R(w_s)$. It was found in
\cite{7,8} that
\begin{equation}
\langle 0|\bar q(0)q(0)|0\rangle_s =\frac{C}{R^2(w_s)\rho}\,,
\label{67b}
\end{equation}
with $C=25.0$. Thus we can study dependence of the solution of the sum rules equations on the fraction of the small size instantons $w_s$ for several values of the scalar condensate $a=-(2\pi)^2 \langle 0|\bar q(0)q(0)|0\rangle$ and of the effective size of small instantons $\rho$.

Note that at $\rho=0.33$~fm and $R(1)=1$~fm the scalar condensate $a=0.58\,{\rm GeV}^3$ (at the conventional normalization point $\mu=0.5$ GeV) \cite{7,8}.
This enables us to find the dependence on $w_s$ at any values of $a$ and $\rho$.

The results for $\rho=0.33$~fm are presented in Table 1 and in Fig.~5.  One can see that at several reasonable values of the quark condensate the sum rules have a physical solution for $w_s$ which does not exceed certain value
$w_0$. At $w_s=w_0 \approx 0.67$ the solutions jump to unphysical ones with a smaller value of the nucleon mass and the dominative contribution of the continuum \cite{11a}. At $w_s$ about $0.6$ the nucleon mass is close to the physical value.

In Table 1 and in Fig.~5 we present the results for four values of the scalar condensate $a$ corresponding to
$\rho=0.33$~fm and the distances between the small instantons $R=1.3$~fm, $R=1.2$~fm,  $R=1.1$~fm and $R=1.0$~fm at $w_s=0.6$.
The distances $R=1.3$~fm and $R=1.2$~fm correspond to the values $a=0.58\,{\rm GeV}^3$ and $a=0.67\,{\rm GeV}^3$, i.e., to the values of the scalar condensate $\langle 0|\bar q(0)q(0)|0\rangle$ equal to $(-244\,{\rm MeV})^3$ and  $(-257\,{\rm  MeV})^3$, close to conventional values.
The distances $R=1.1$~fm and $R=1.0$~fm correspond to  $a=0.80\,{\rm GeV}^3$ and $a=0.96\,{\rm GeV}^3$, i.e., to somewhat larger values of $\langle 0|\bar q(0)q(0)|0\rangle$ equal to $(-273\,{\rm MeV})^3$ and to less realistic $(-290\,{\rm MeV})^3$.
The consistency of the LHS and RHS of the sum rules is illustrated by Fig.~6.

As we said earlier, the pole-to-continuum ratio
\begin{equation}
r_i(M^2)={\cal F}_i^{p}(M^2)/{\cal F}_i^{c}(M^2); \quad i=q,I
\label{80}
\end{equation}
of the two contributions to the RHS of Eq.~(\ref{3g}) is the characteristic of validity of the "pole + continuum" model for the spectrum of polarization operator - Eqs.~(\ref{3g},\ref {3h}). For larger values of $r_i(M^2)$ the model is justified better. The values of the ratio are presented in Table 2 for  $\rho=0.33$ ~fm, $w_s=0.60$. We took two cases for illustration. For $a=0.57\,{\rm GeV}^3$ solution is
\begin{equation}
m_N = 1008\,{\rm MeV}; \quad \lambda^2=1.2\,{\rm GeV}^6; \quad W^2=2.0 \mbox{GeV}^2\,.
\label{80a}
\end{equation}
The pole-to-continuum ratio decreases with the value of $M^2$ - see Table 2.
Although
the SR equations can be solved with good accuracy in the broad interval of the values of the Borel mass (see Table 3),
the pole-to-continuum ratio becomes unacceptably small for $M^2>1.4\,{\rm GeV}^2$.
Thus in this case we stay in the traditional duality interval determined by Eq.~(\ref{230}).

For the condensate $a=0.96~$GeV$^3$, corresponding to $R(0.6)=1$~fm the solution is
\begin{equation}
m_N = 1147\,{\rm MeV}; \quad \lambda^2=2.8\,{\rm GeV}^6; \quad W^2=2.9~{\rm GeV}^2\,.
\label{80z}
\end{equation}
Here the SR equations also can be solved with good accuracy in the large interval of the values of the Borel mass - see Table 3.
One can see that both $r_q$ and $r_I$ decrease while value of $M^2$ increases.
In this case the pole-to-continuum ratio is much larger than it was for the smaller values of the condensate.
Thus the interval of the values of $M^2$ where the SR equations can be solved becomes larger.

We also fix the value of $R=1.3$~fm and  trace the dependence of the solutions on variation of $\rho$. In Table 4 we present the results for $\rho=0.25$~fm ($a=0.76\,{\rm GeV}^3$; $\langle 0|\bar q(0)q(0)|0\rangle=(-268\,{\rm MeV})^3)$ and $\rho=0.40$~fm
($a = 0.48\,{\rm GeV}^3$; $\langle 0|\bar q(0)q(0)|0\rangle = (-230\,{\rm MeV})^3)$.
They are shown in Fig.~7. The situation is similar to the previous case when we changed $R$. However, at $\rho=0.40$~fm the jump to the unphysical solution takes place at a larger value
$w_s \approx 0.75$.

For $w_s=0.65$ the function $K(M^2)$ determined by Eq.~(\ref{80c}) is approximated by the series in the RHS
of Eq.~(\ref{80d}) with parameters
\begin{equation}
C_1=-0.80\mbox{ GeV}^2; \quad C_2=0.35\mbox{ GeV}^4\,,
\label{90}
\end{equation}
and thus $c_1=-0.10\,{\rm GeV}^2$, $c_2=0.011\,{\rm GeV}^4$.

\section{Summary}

We calculated the polarization operator of the nucleon current in the instanton medium which we assumed to be a composition of the instantons of large and small sizes. The instantons of large size $\rho \gg  (1\,{\rm GeV})^{-1}$ manifest themselves in terms of the local scalar quark condensate. The quark propagator in the field of the small size instantons contained the zero mode chirality flipping part proportional to effective quark mass $m(p)$ and the nonzero mode part approximated by the propagator of the free massless quark \cite{7,8}. The zero-mode part can be expressed in terms of the nonlocal scalar condensate.

We solved the sum rules equations and trace the dependence on the solution on the fraction of the small size instantons $w_s$. We demonstrated that at $w_s \leq 0.6-0.7$ the sum rules have a solution with a reasonable value of the nucleon mass.
At $w_s \approx 2/3$ the value of the nucleon mass is very close to the physical one. The numerical values vary
slightly with variation of the actual values of the size of small instantons and of the distance between them.
Finally at the values of the scalar condensate close to the conventional value $(-250\,{\rm MeV})^3$
\begin{equation}
m_N \approx 1\mbox{ GeV}; \quad \lambda^2 \approx 1\mbox{ GeV}^6; \quad W^2 \approx 2\mbox{ GeV}^2\,.
\label{101}
\end{equation}
At larger values of $w_s$ the sum rules have only an unphysical solution with a strong domination of the continuum contribution over that of the nucleon pole and with a small value of the nucleon mass.

The solution (\ref{101}) was found for $\rho=0.33$~fm, while $R=1.2-1.3$~fm. It is valid also for $r \approx 1.3$~fm while $\rho \sim 0.25-0.40$~fm. At larger values of the quark condensate the values of the nucleon residue and of the continuum threshold increase, reaching the values $\lambda^2 \approx 3\,{\rm GeV}^6$ and $W^2\approx 3\,{\rm GeV}^2$ at $\langle 0|\bar q(0)q(0)|0\rangle=(-290\,{\rm MeV})^3$.

Comparing to the SR in the condensate presentation, we included the nonlocality of the scalar condensate.
Also, inclusion of instantons strongly diminished the role of the contribution corresponding to four quark condensate in the condensate language.

The consistency between the LHS and RHS of the sum rules appeared to be much better than in the sum rules
in terms of local condensates, where the value of "$\chi^2$ per point" was of the order $10^{-1}$ \cite{10}. At larger values of the scalar condensate the domination of the contribution of the pole over that of the continuum becomes more pronounced. Also, the duality interval becomes larger than that defined by Eq.~(\ref{230}) due to the shift of the upper limit.

We demonstrated that the contribution of nonlocality of the  scalar condensate can be approximated by two additional terms of $1/M^2$ series. This corresponds to approximation of the dependence of the nonlocal quark condensate $f(x^2)=\langle 0|\bar q(x)q(0)|0\rangle$ on $x^2$ by a polynomial of the second order. At $x^2=1\,{\rm GeV}^{-2}$ (Euclidean metric) we found $f(x^2)-f(0)=tf(0)$ with $t=-0.14$ for $w_s=1$ and $t=-0.09$ for $w_s=0.65$. More complicated calculations in framework of the instanton liquid model \cite{Shur} provided $t \approx -0.1$ for $x^2=1\,{\rm GeV}^{-2}$.
The parameter $m_0^2$ defined by Eq.~(\ref{3dd}) determines the lowest order term of the Taylor series of the condensate $f(x^2)$. Its value was estimated in the nucleon QCD sum rules analysis as providing the best fit of the two sides of the sum rules.
The result of \cite{I1} is $m_0^2 \approx 0.8\,{\rm GeV}^2$ leading to $t \approx 0.2$, while the value  $m_0^2 \approx 0.2\,{\rm GeV}^2$ providing $t \approx-0.05$ was obtained in \cite{1a}.

Note that these are to large extent the preliminary results. In next steps we plan to include interactions between the quarks composing the polarization operator, i.e., we must take into account the radiative corrections. They are the same as in the condensate presentation for the structure $\Pi^q$. However, additional work is required to find these corrections for the chirality flipping structure $\Pi^I$. Another point is the interpretation of the condensate $(1-w_s) \langle 0|\bar qq|0\rangle$.
In the present paper we attributed it to the condensates of large size. A more general analysis is required.
The results will be published elsewhere.

The attempts to include the instantons into the SR were made in \cite{12,13,14}. Actually the authors were interested in description of the quark correlations in the polarization operator due to interaction with the same instanton.  In other words the authors considered the instanton correction to the Green function of {\em two} quarks but neglect the instanton effect in the propagation of one quark where just the "mean field" scalar condensate $\langle0|\bar q(0)q(0)|0\rangle$ was included.\\

We thank A.E. Dorokhov, N. I. Kochelev and especially V. Yu. Petrov and M. G. Ryskin for stimulating discussions.
We also acknowledge the partial support by the RFBR grant 12-02-00158.

\def\thesection{Appendix \Alph{section}}
\def\theequation{\Alph{section}.\arabic{equation}}
\setcounter{section}{0}

%  A.
\section{}
\setcounter{equation}{0}

%\section*{Appendix}

In order to calculate the integral on the RHS of Eq.~(\ref{49}) we present
\begin{equation}
\ln{(Q-p)^2}=-\int_0^{\infty}\frac{dy}{(Q-p)^2+y}\,.
\label{A2}
\end{equation}
Here and below we omit the polynomials in $Q^2$ since they will be eliminated by the Borel transform.
Now we can write
\begin{equation}
X_s=-\frac{3i}{\pi^2}\int\frac{d^4p}{(2\pi)^4} \frac{{\cal A}}{p^2(p^2+\eta^2)^3} \int_0^{\infty}\frac{dy\,y}{(Q-p)^2+y}\,.
\label{A3}
\end{equation}
One can check that it is possible to present
\begin{equation}
\frac{1}{p^2(p^2+\eta^2)^3}=3\int_0^{1}\frac{dx\,x^2}{(p^2+\eta^2x)^4}\,,
\label{A4}
\end{equation}
and thus
\begin{equation}
X_s=-3\int_0^{1}dx\,x^2\Psi(\eta^2x)\,,
\label{A5}
\end{equation}
where
\begin{equation}
\Psi(\mu^2)=\frac{3{\cal A}}{\pi^2}\int_0^{\infty}dy\,y\Phi(\mu^2,y); \quad \Phi(\mu^2,y)=\int \frac{d^4p}{(2\pi)^4} \frac{1}{(p^2+\mu^2)^4}\cdot \frac{1}{(Q-p)^2+y}\,.
\label{A6}
\end{equation}
Carrying out the integration over the angular variables we find
\begin{equation}
\Phi(\mu^2,y)=\frac{1}{48\pi^2}\int_0^1\frac{dt(1-t)^3}{(ty+\mu^2(1-t)+t(1-t)Q^2)^3}=\int_0^1\frac{dt(1-t)^3}{t^3(y+\kappa)^3}; \quad \kappa=Q^2(1-t)+\frac{\mu^2(1-t)}{t^3}\,.
\label{A7}
\end{equation}
Carrying out integration over $y$ we obtain
\begin{equation}
\Psi(\mu^2)=\frac{{\cal A}}{32\pi^4}\int_1^{\infty}{du}\left(1-\frac{1}{u}\right)^2\frac{u}{Q^2+\mu^2u}\,.
\label{A8}
\end{equation}
The divergence on the upper limit is not important, since this contribution will be eliminated by the Borel transform.
Returning to Eq.~(\ref{A5}) we can write it as
\begin{equation}
X_s=\frac{3{\cal A}}{32\pi^4}\int_0^1dx\,x^2\int_1^{\infty}{du} \left(1-\frac{1}{u}\right)^2 \frac{u}{Q^2+\eta^2ux}\,.
\label{A9}
\end{equation}
Now integration can be carried out easily, providing
\begin{equation}
X_s=\frac{3{\cal A}}{32\pi^4}\Big[\frac{Q^4}{\eta^6}\ln{\frac{Q^2+\eta^2}{Q^2}}+(\frac{3Q^2}{\eta^4}+\frac{3}{\eta^2}+
\frac{1}{Q^2})\ln{\frac{Q^2+\eta^2}{\eta^2}}\Big]\,.
\label{A10}
\end{equation}
After the Borel transform we come to Eq.~({\ref{55}).
Note that carrying ont the Borel transform of the RHS of Eq.~(\ref{A9}) we come to a compact expression
\begin{equation}
{\cal B}X_s=\frac{3{\cal A}}{32\pi^4}\int_0^1dx\,x^2\int_1^{\infty}du\,u \left(1-\frac{1}{u}\right)^2 exp(-\eta^2xu/M^2)\,.
\label{A11}
\end{equation}

{}

%\end{document}
\clearpage

\newpage
\begin{table}
\caption{Solutions of the sum rules equations for $\rho=0.33$ fm.}
\begin{center}
\begin{tabular}
{|c|c|c|c|c|c|} \hline
$a$,\, GeV$^3$&$w_s$&$m_N,$\,MeV&$\lambda^2$,\,GeV$^6$&$W^2$,\,GeV$^2$&$\chi^2_N$\\
\hline
      &0.30&1452 & 8.7&6.6&3.7(-2)\\
0.96 &0.60&1147 & 2.8 &2.9&4.0(-2)\\
      &0.66&1052 & 1.9 &2.3&3.9(-2)\\
      &0.67&822  & 0.86&1.4 &2.1(-2)\\

\hline

      &0.30&1396 &6.2 &4.9 &1.7(-2)\\
0.80 &0.60&1103 &2.0  &2.6 &2.3(-2)\\
      &0.67&994  &1.2  &2.0 &2.2(-2)\\
      &0.68&800  &0.60 &1.3 &1.2(-2)\\

\hline

      &0.30 &1330& 4.3  &4.0 &8.3(-3)\\
0.67 &0.60 &1052& 1.4  &2.2 &1.4(-2)\\
      &0.67 &948 & 0.83 &1.7 &1.3(-2)\\
      &0.68 &771 & 0.41 &1.1 &5.9(-3)\\

\hline
      &0.30& 1269& 3.0 & 3.4 &4.2(-2)\\
0.57 &0.60& 1002& 0.95& 1.9 &8.9(-3)\\
      &0.67& 897 & 0.57& 1.5 &8.3(-3)\\
      &0.68& 747 & 0.30& 1.0 &3.0(-3)\\

\hline

\end{tabular} \end{center}
\end{table}

\begin{table}
\caption{Pole-to-continuum ratio $r(M^2)$ for solutions of the sum rules at $\rho=0.33$ fm
for $a= 0.58\,{\rm GeV}^3$ and $a= 0.96\,{\rm GeV}^3$; $w_s=0.60$.}

\begin{center}
\begin{tabular}{|c|c|c|c|} \hline
$a,\,{\rm GeV}^3$ & $M^2,\,{\rm GeV}^2$ & $r_q(M^2)$ & $r_I(M^2)$ \\
\hline

&0.8&1.25&1.84\\

0.58 &1.0&0.69&1.08 \\

&1.2&0.43&0.72 \\

&1.4&0.29&0.52\\

\hline
      &0.8& 4.69& 5.85\\
0.96 &1.0& 2.30& 2.99 \\
      &1.2& 1.34& 1.82 \\
      &1.4& 0.86& 1.23\\

\hline

\end{tabular} \end{center}
\end{table}

\begin{table}
\caption{Solutions of the sum rules equations in various intervals of the values of the Borel mass.
The values of parameters are the same as in Table 2.}
\begin{center}
\begin{tabular}{|c|c|c|c|c|c|} \hline
$a,\,{\rm GeV}^3$ & $M^2,\,{\rm GeV}^2$ & $m_N,\,{\rm MeV}$ & $\lambda^2,\,{\rm GeV}^6$ & $W^2,\,{\rm GeV}^2$ & $\chi^2_N$\\
\hline

      &0.8 - 1.4&1006&0.98&1.96&9.3(-3) \\

0.58 &0.8 - 1.6&1017&1.01&1.99&1.2(-2)\\

      &0.8 - 1.8&1026&1.04&2.01&1.5(-2)\\
\hline

      &0.8 - 1.4& 1147& 2.83& 2.93 & 4.0(-2) \\
0.96 &0.8 - 1.6& 1173& 3.03& 3.02 & 5.1(-2)\\
      &0.8 - 1.8& 1193& 3.20& 3.08 & 6.0(-2)\\
\hline
\end{tabular} \end{center}

\end{table}

\begin{table}
\caption{Solutions of the sum rules equations for $R \approx 1.3$ fm.}

\begin{center}
\begin{tabular}
{|c|c|c|c|c|c|} \hline
$a,\,{\rm GeV}^3$ & $w_s$ & $m_N$,~MeV & $\lambda^2,\,{\rm GeV}^6$ & $W^2,\,{\rm GeV}^2$ &
$\chi^2_N$\\
\hline

      &0.60 & 1094& 1.52 & 2.31 &3.8(-3)\\
0.77 &0.70 & 879 & 0.55 & 1.38 &1.5(-3)\\
      \hline
      &0.50 & 1059& 1.20 & 2.21 & 7.8(-3)\\
0.48 & 0.60&979 &0.82   &1.87 & 1.1(-2)\\
      \hline

\end{tabular} \end{center}
\end{table}

\clearpage

\section*{Figure captions}
\noindent
Fig.~1. The set of the diagrams for the lowest OPE terms of the nucleon sum rules.  Wavy lines are for the nucleon current, solid lines stand for the quarks, dashed lines denote the gluons. The circles stand for the quark and gluon condensates.\\

\noindent
Fig.~2. The set of the diagrams for the  quarks in the fields of instantons.
 Dark and dashed blobs on the quark lines stand for the small size and large size instantons.\\

\noindent
Fig.~3.~$a$: The function $F(\beta)$ determined by Eq.~(41).\\
$b$: Dependence of the  functions $F(\eta^2/M^2)$ for $\eta^2=1.26\,{\rm GeV}^2$, corresponding to the size $\rho=0.33$~ fm.\\
\noindent
Fig.~4. Approximation of the function $K(M^2)$ defined by Eq.~(\ref{80c}) (solid line) by the series on the RHS of Eq.~(\ref{80d}) with parameters determined by Eq.~(\ref{82yyy}) (dotted).

\noindent
Fig.~5. Dependence of the solution of the sum rules equations on the value of $w_s$ at $\rho=0.33$~fm. Fig.~5$a$ is for the nucleon mass, Fig.~5$b$ demonstrates the $w_s$ dependence of $\lambda^2$, Fig.~5$c$ is for $W^2$. The solid, dashed, dotted and dashed-dotted lines are for the values of the scalar condensate
$a=0.58\,{\rm GeV}^3$, $a=0.67\,{\rm GeV}^3$,  $a=0.80\,{\rm GeV}^3$, and $a=0.96\,{\rm GeV}^3$.

Fig.~6. Consistency of the $LHS$ and $RHS$ of the sum rules for the case $a=0.58\,{\rm GeV}^3$, $w_s=0.60$.
The solid and dashed lines show the RHS-to-LHS ratios for the SR for chirality conserving and chirality flipping equations, correspondingly. \\

\noindent
Fig.~7. Dependence of the solution of the sum rules equations on the value of $w_s$ at $R \approx 1.3$~fm for Fig.~7$a$ is for the nucleon mass, Fig.~7$b$ demonstrates the $w_s$ dependence of $\lambda^2$, Fig.~7$c$ is for $W^2$. The solid, dashed and dotted  curves are for the values of the scalar condensate $a=0.58\,{\rm GeV}^3$, $a=0.48\,{\rm GeV}^3$ and $a=0.77\,{\rm GeV}^3$,  correspondingly.\\

\clearpage

\newpage
\begin{figure}%1
\centerline{\epsfig{file=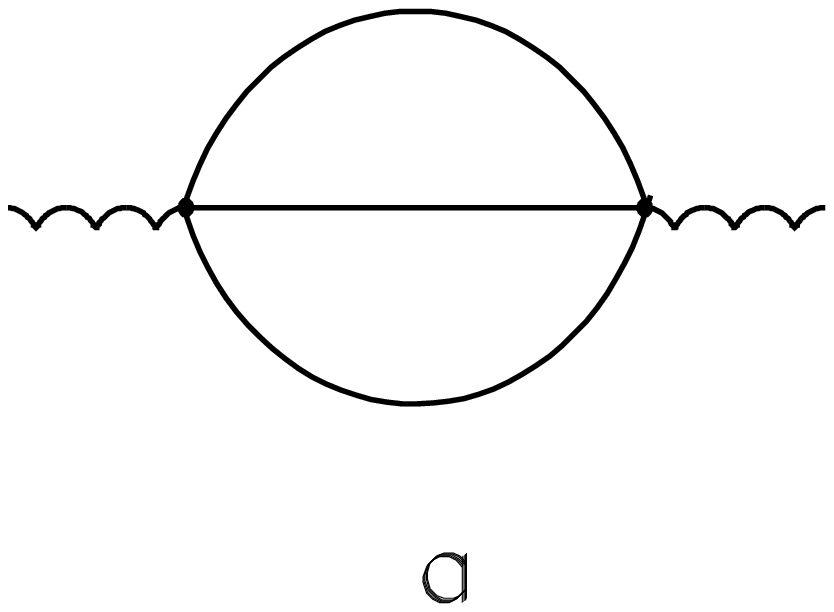,width=5cm}
\epsfig{file=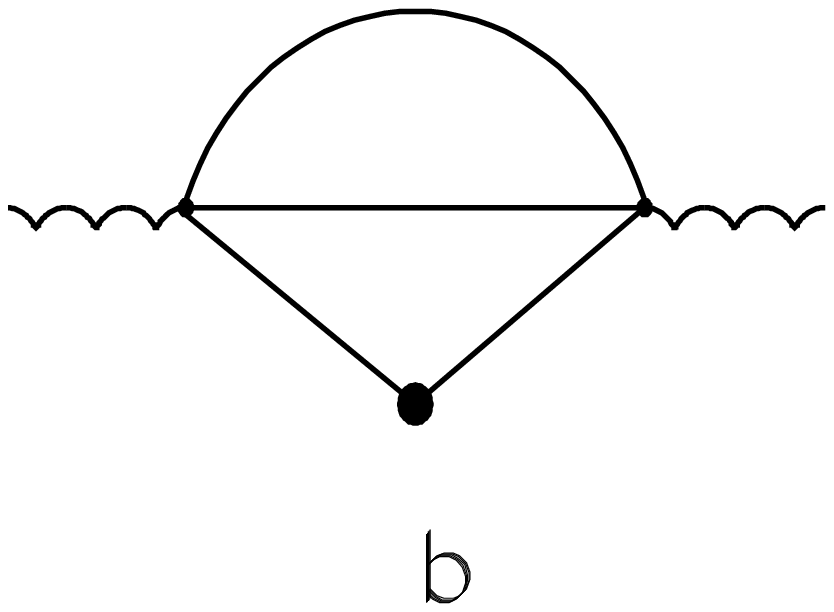,width=5cm}}
\centerline{\epsfig{file=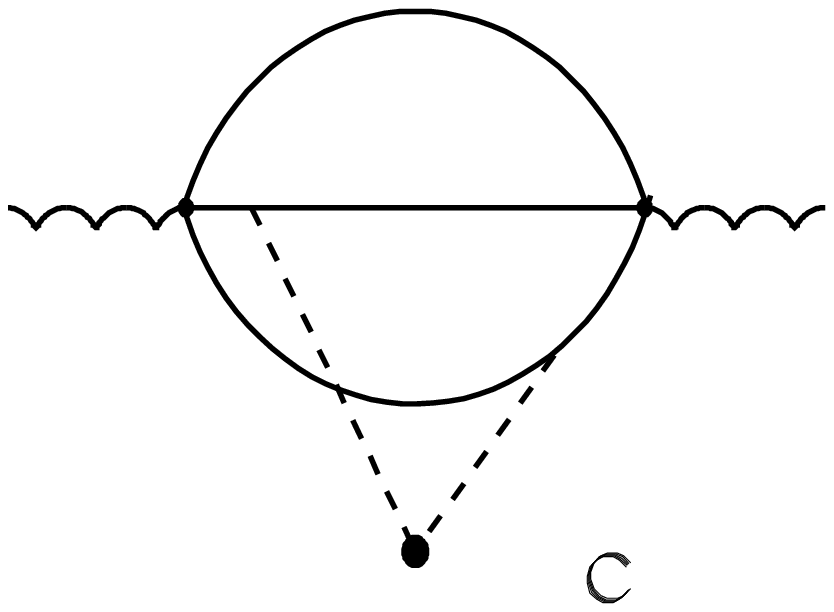,width=5cm}
\epsfig{file=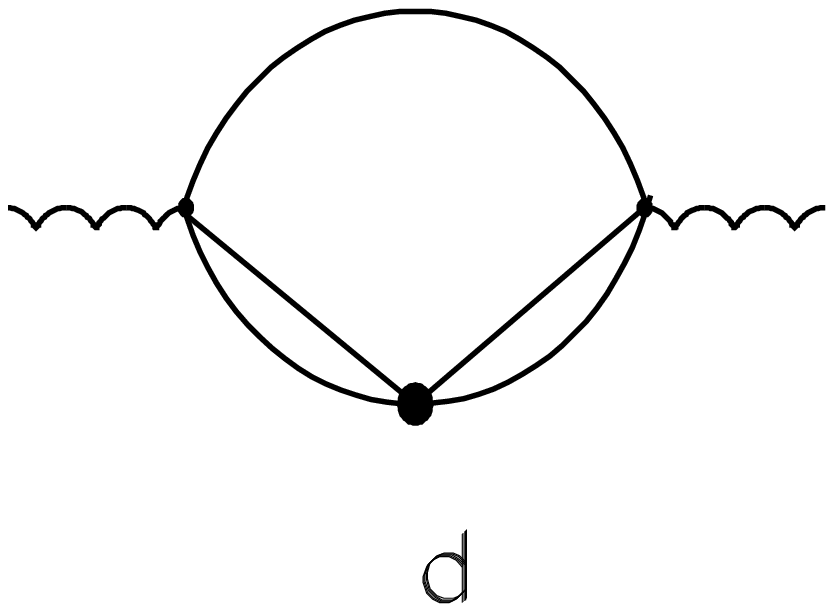,width=5cm}}
\caption{}
\end{figure}

\begin{figure}%2
\centerline{\epsfig{file=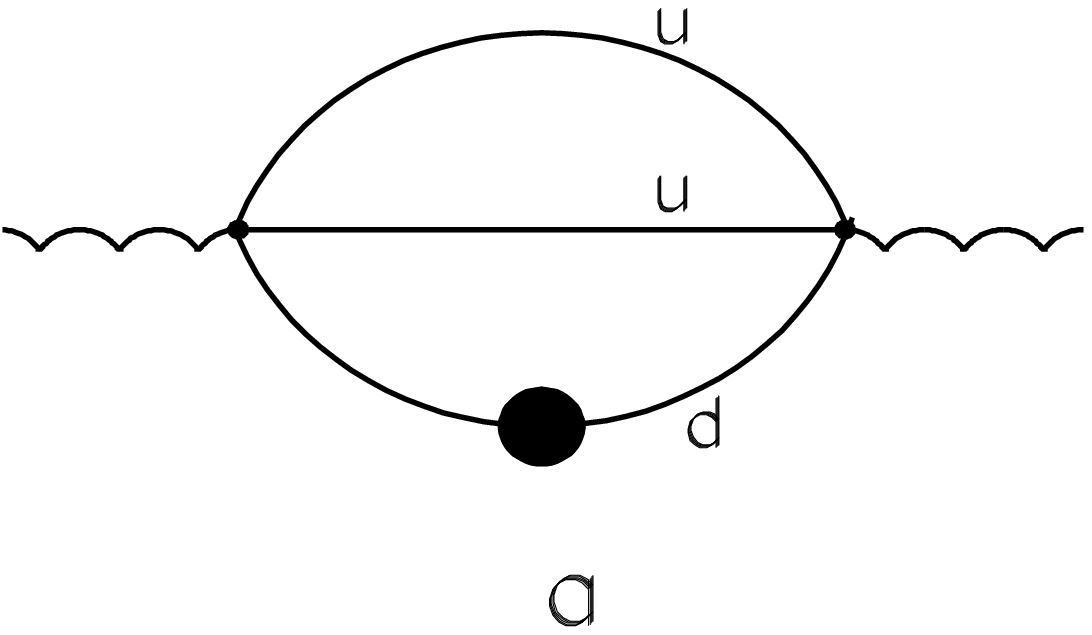,width=5cm}\epsfig{file=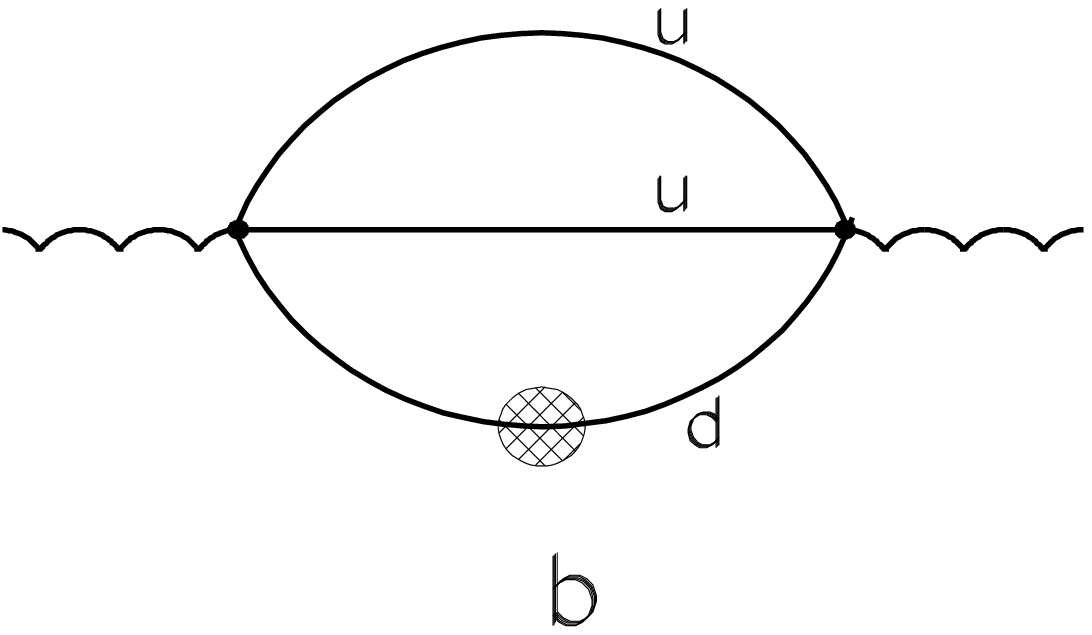,width=5cm}}
\centerline{\epsfig{file=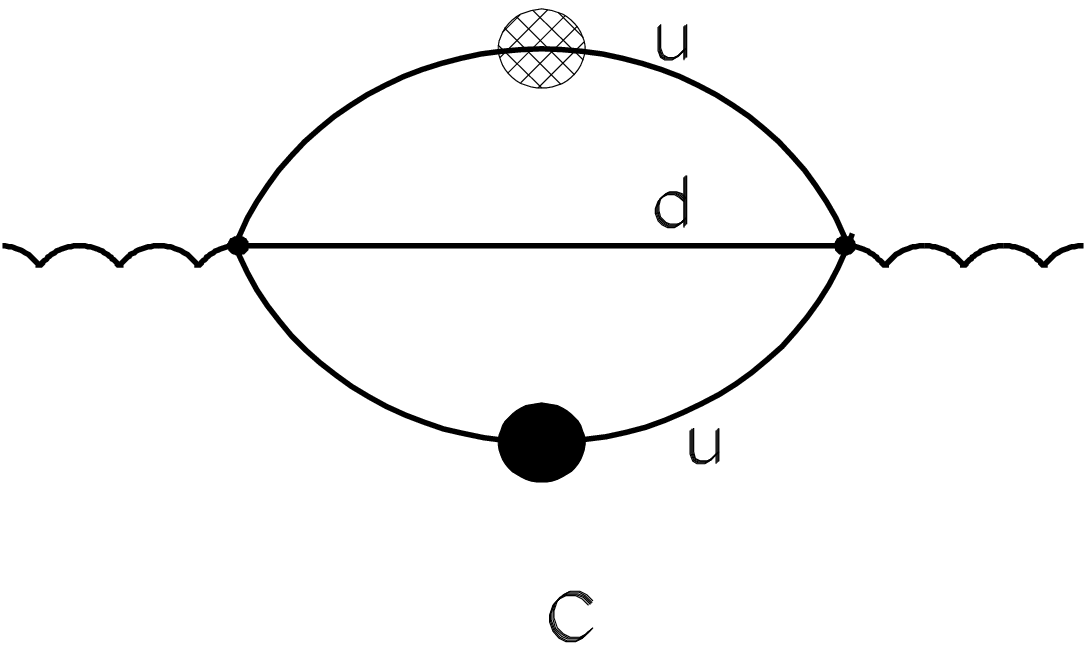,width=5cm}\epsfig{file=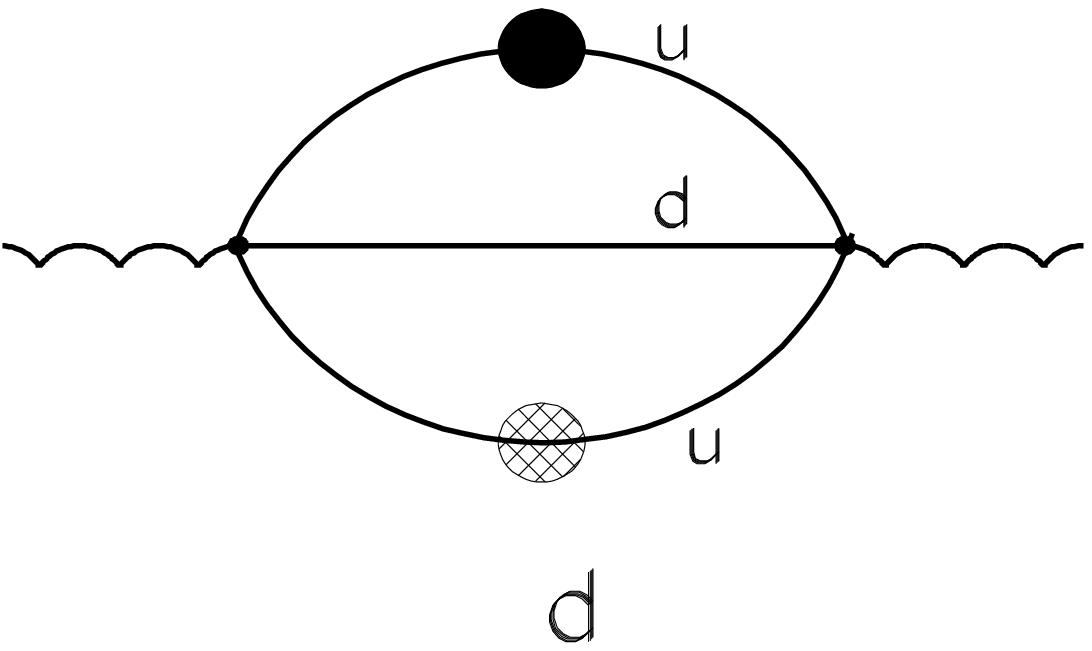,width=5cm}}
\caption{}
\end{figure}

\begin{figure}[ht] %Fig.3
\centerline{\epsfig{file=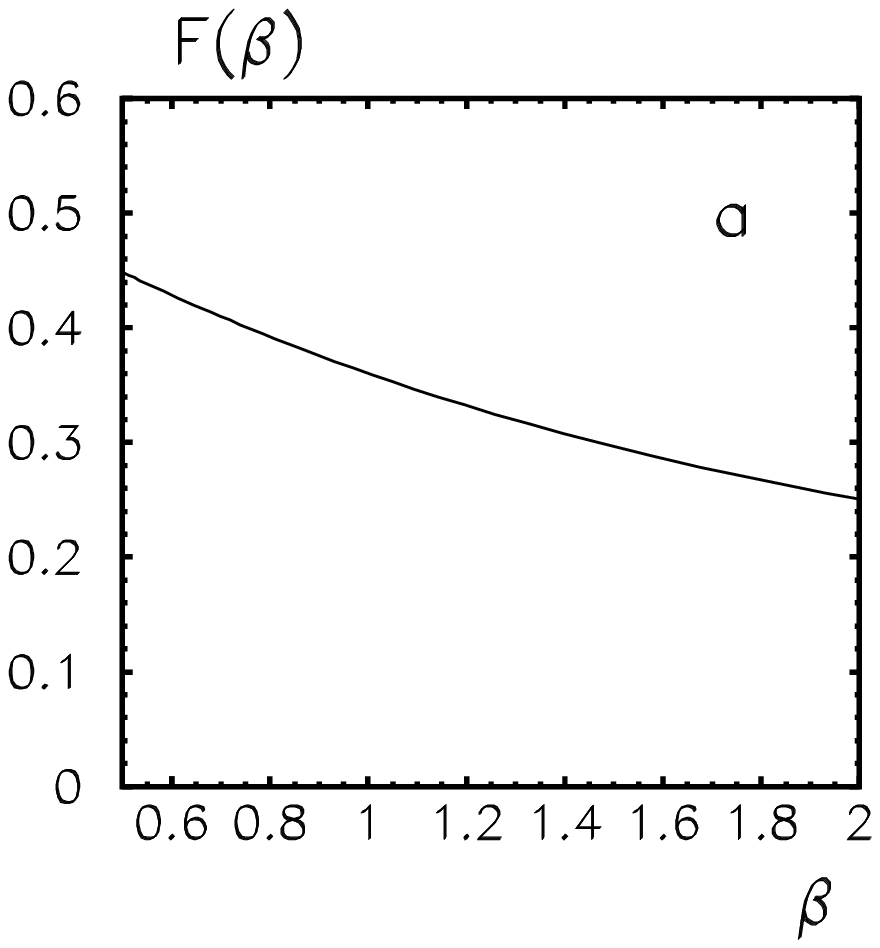,width=6.0cm} \epsfig{file=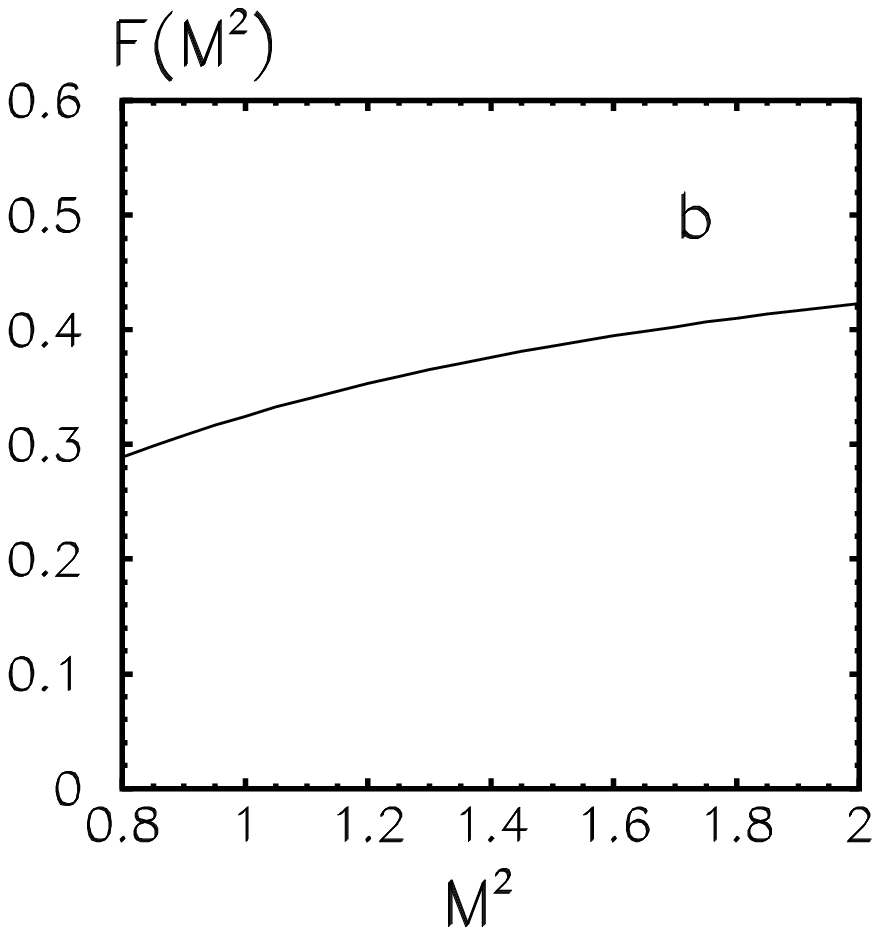,width=6.0cm}}
 \caption{}\end{figure}

\begin{figure}%4
\centerline{\epsfig{file=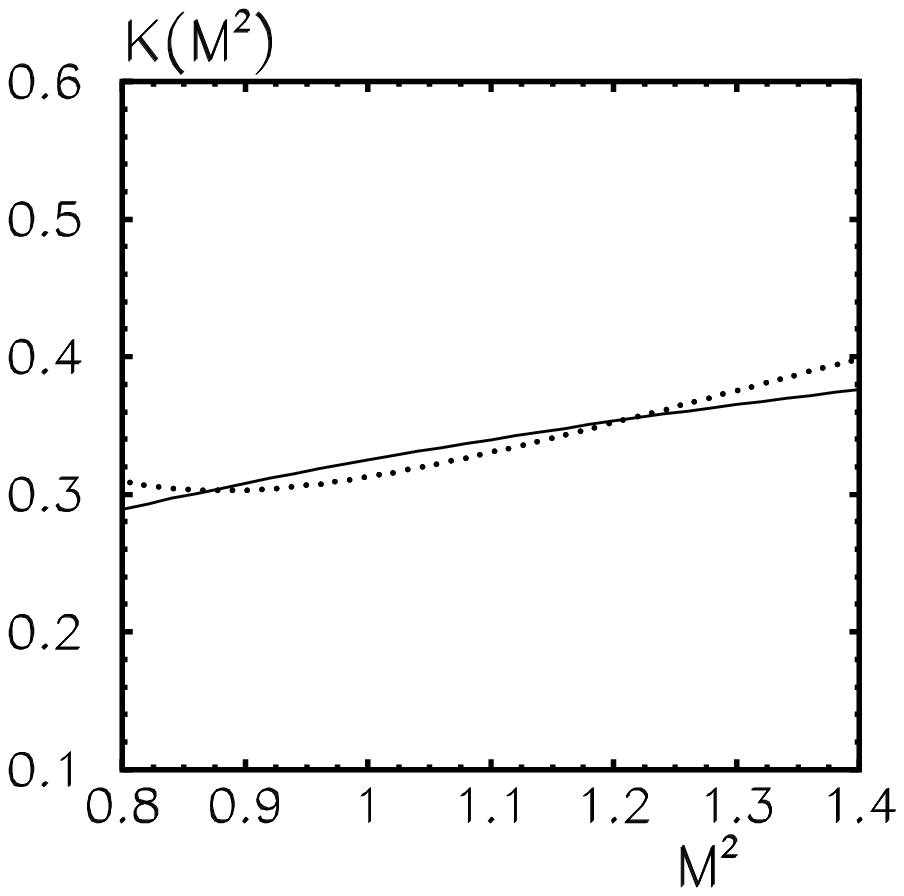,width=6cm}}
\caption{}
\end{figure}

\begin{figure}%5
\centerline{\epsfig{file=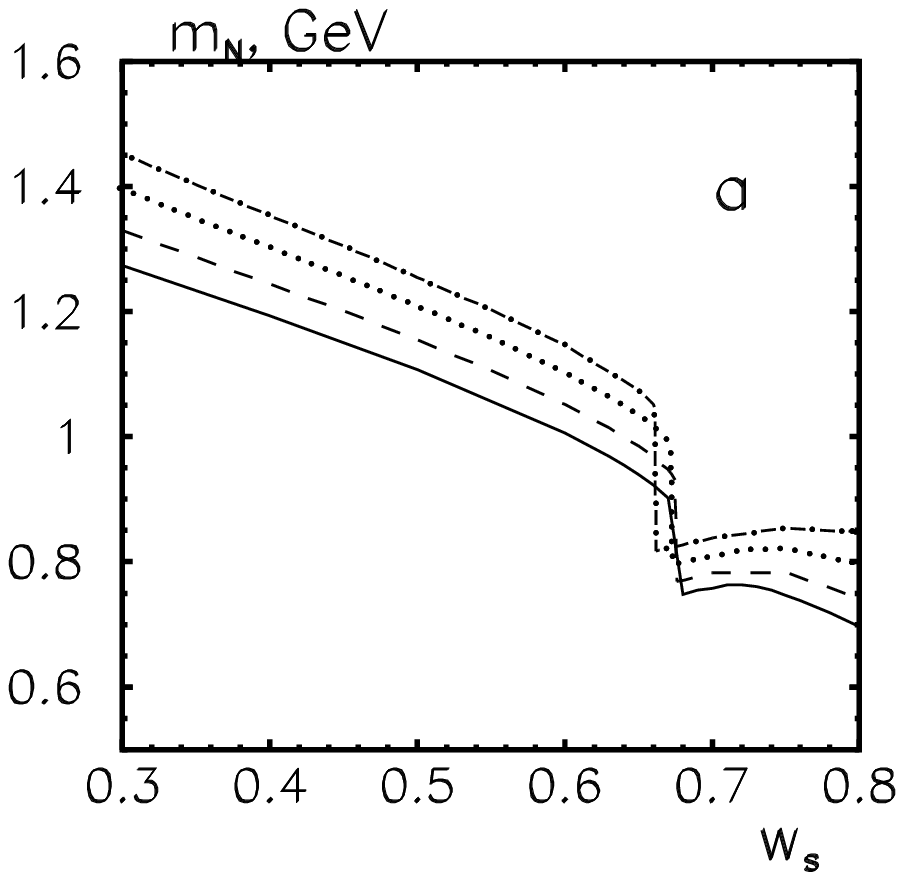,width=6cm}\epsfig{file=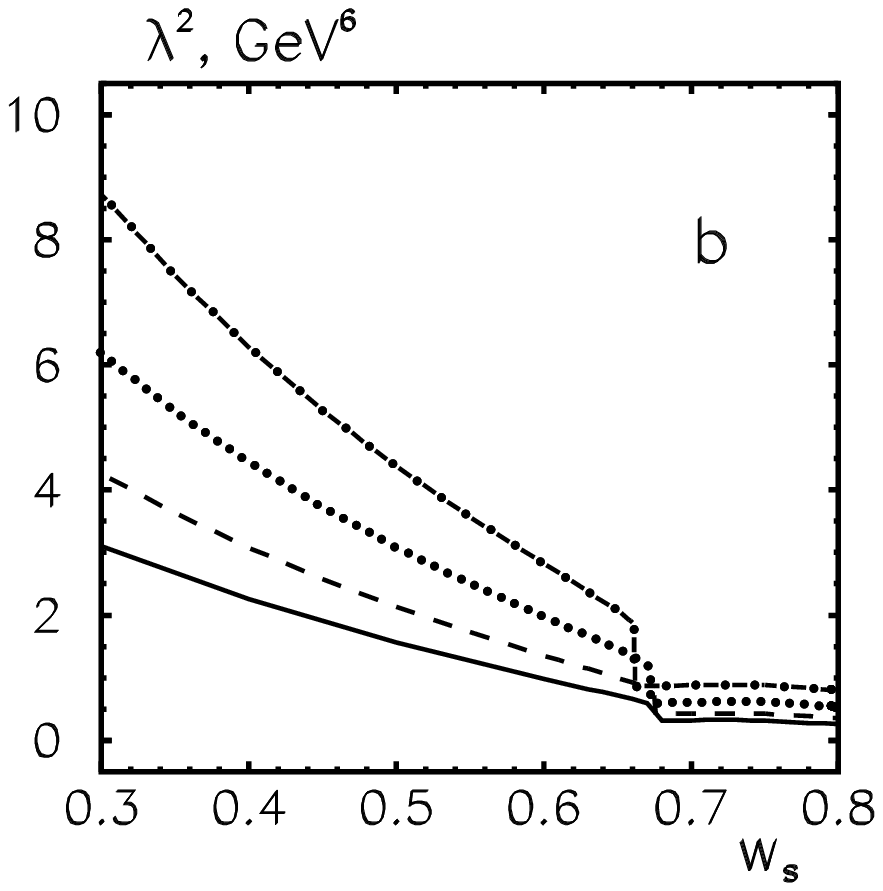,width=6cm}}
\centerline{\epsfig{file=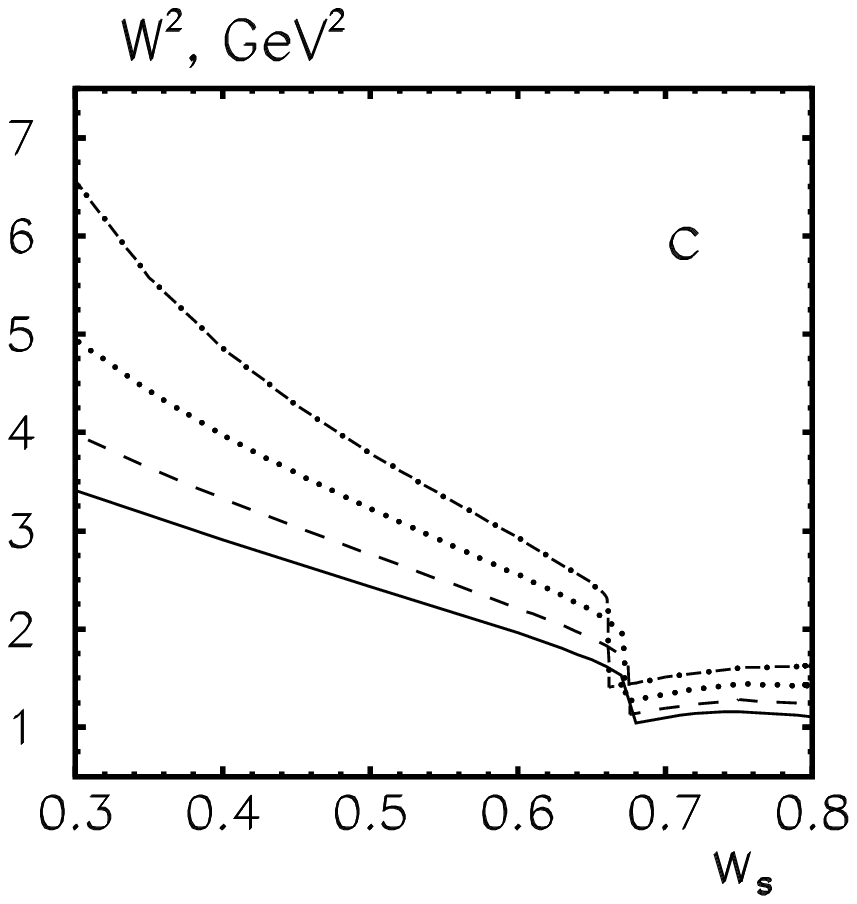,width=6cm}}
\caption{}
\end{figure}

\begin{figure}%6
\centerline{\epsfig{file=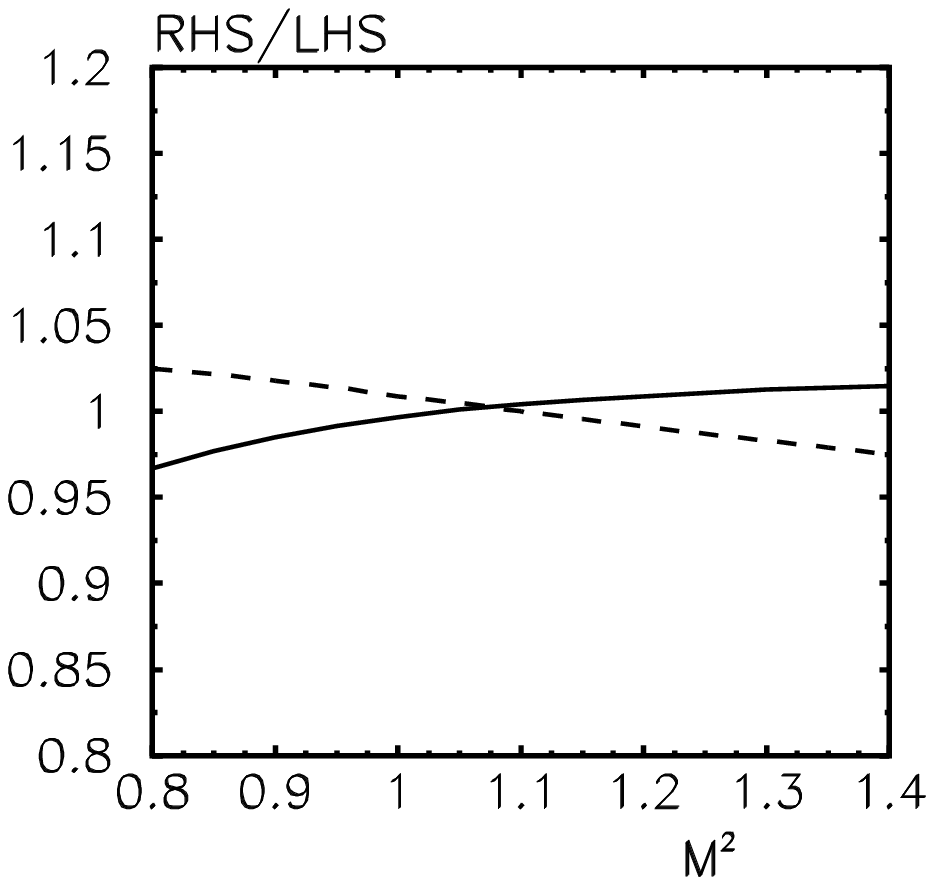,width=6cm}}
\caption{}
\end{figure}

\begin{figure}%7
\centerline{\epsfig{file=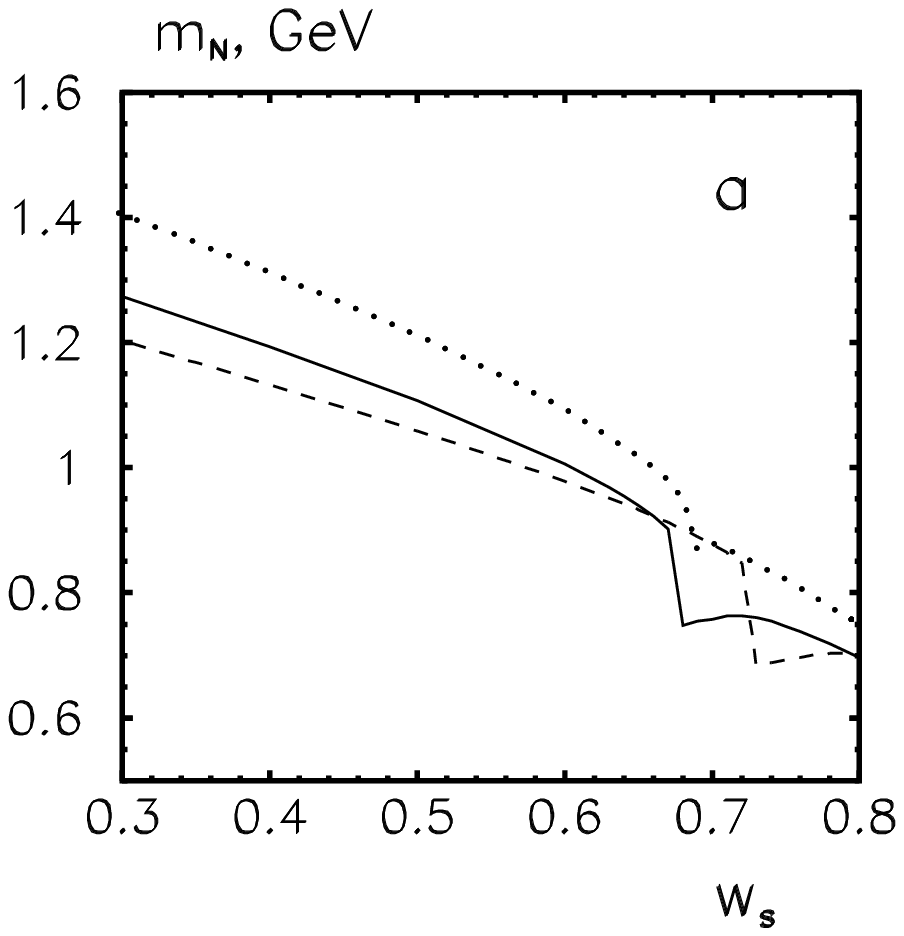,width=6cm}\epsfig{file=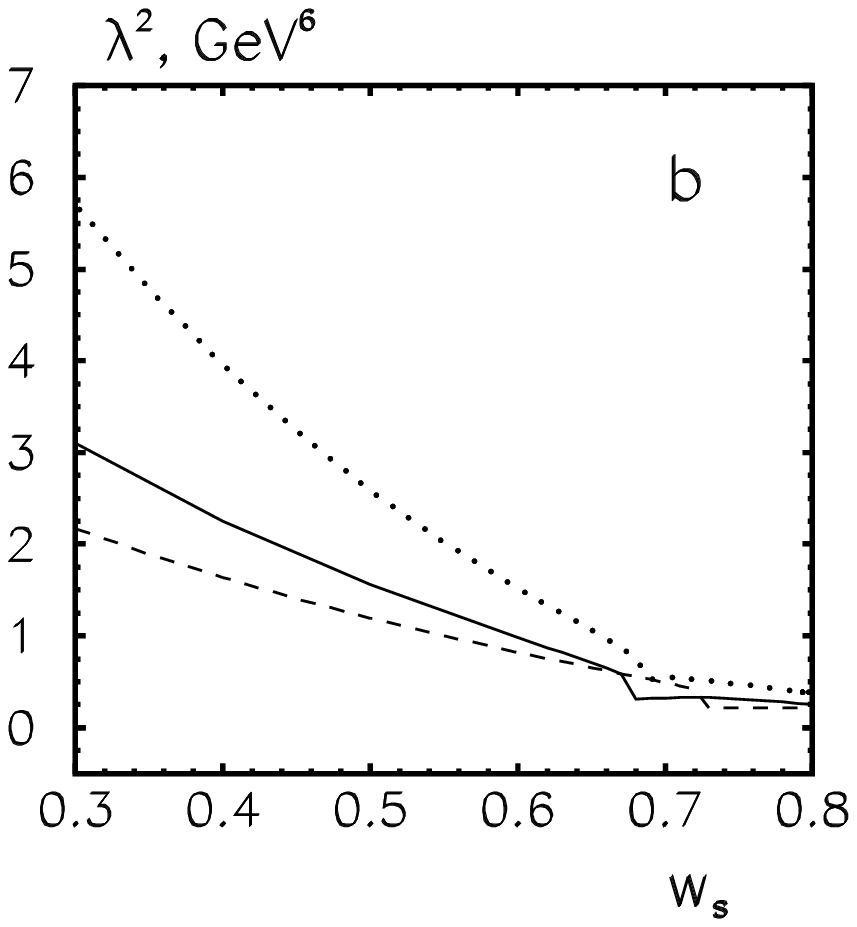,width=6cm}}
\centerline{\epsfig{file=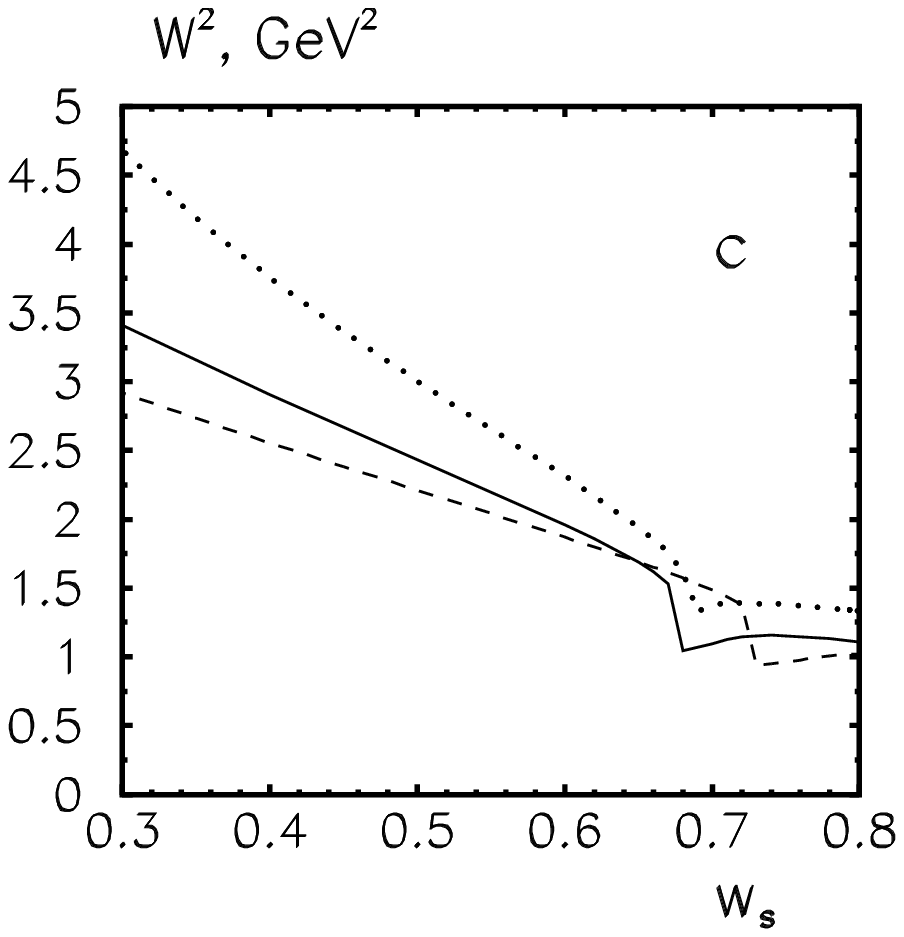,width=6cm}}
\caption{}
\end{figure}


\clearpage

\newpage

\begin{thebibliography}{}

\bibitem{1}M.A. Shifman, A.I. Vainshtein and V.I.~Zakharov, Nucl.
Phys. ~{\bf B147}, 385; 448; 519 (1979).

\bibitem{2} B.L. Ioffe, Nucl. Phys. ~{\bf B 188}, 317 (1981); {\bf B 191}, 591(E) (1981).

\bibitem{3} B.L. Ioffe, L.N. Lipatov and V.S.~Fadin, {\em Quantum
Chromodynamics} (Campridge Univ. Press, 2010).
\bibitem{4} K. G. Wilson, Phys. Rev. {\bf 179}, 1499 (1969).

\bibitem{5}E. V. Shuryak, {\em The QCD Vacuum, Hadrons and the Superdense Matter}, World Scientific Pub. Co, Singapore, 1988.
\bibitem{6} A. Ringwald, F. Schrempp, Phys. Lett. {\bf B459}, 249 (1999).
\bibitem{7}  D. I. Dyakonov and V. Yu. Petrov, Sov. Phys. ZhETP, {\bf 89}, 361 (1985).
\bibitem{8}  D. I. Dyakonov and V. Yu. Petrov, Nucl. Phys. B {\bf 272}, 457 (1986).

\bibitem{9} B.L. Ioffe, Z. Phys. C {\bf18}, 67 (1983).
\bibitem {I1} B. L. Ioffe and A. V. Smilga, Nucl.Phys.B {\bf 232},109 (1984).
\bibitem{10} V.~A.~Sadovnikova, E.~G.~Drukarev and M.~G.~Ryskin, Phys.
Rev. D~{\bf 72}, 114015 (2005).
\bibitem{M1} M. G. Ryskin, E. G. Drukarev, V. A. Sadovnikova, ArXiv: 1405.6828 [hep-ph].
\bibitem{11a} E.~G.~Drukarev, M.~G.~Ryskin, and V.~A.~Sadovnikova, Phys.
Rev. D~{\bf 80}, 014008 (2009).
\bibitem{Shur} E.V. Shuryak, Nucl. Phys. {\bf B328}, 85 (1989).
\bibitem{1a} Y. Chung, H. G. Dosch, M. Kremer, D. Schall, Z. Phys. C {\bf 25}, 151 (1984).
\bibitem{X} M.~G.~Ryskin, E. G. Drukarev, and V.~A.~Sadovnikova, to be published.
\bibitem{12} A. E. Dorokhov, N. I. Kochelev, Z. Phys. C~{\bf 46}, 281
(1990).

\bibitem{13} H. Forkel, M. K. Banerjee, Phys. Rev. Lett. {\bf71}, 484
(1993).
\bibitem{14} Hee-Jung Lee, N. I. Kochelev, V. Vento, Phys. Lett. B~{\bf 610}, 50
(2005).

\end{thebibliography}
\end{document}